\documentclass[runningheads]{llncs}
\usepackage{subcaption}
\usepackage{graphicx}
\usepackage{url}
\usepackage[utf8]{inputenc}
\usepackage[T1]{fontenc}
\usepackage{epstopdf}
\usepackage{soul}
\usepackage{cite}
\usepackage{amsmath}
\begin{document}

\title{Brain Tumor Synthetic Data Generation with Adaptive StyleGANs\thanks{This preprint has not undergone peer review (when applicable) or any post-submission improvements or
corrections. The Version of Record of this contribution is published in [insert volume title],
and is available online at https://doi.org/doi to be updated
. Upon publishing, final version will be available from Springer Nature Publishers (link to be updated). Pre-print of work submitted to 30th AICS conference.}}

%
%
\author{Usama Tariq\inst{1} \and
Rizwan Qureshi\inst{1, 2,3} \and
Anas Zafar\inst{1} \and
Danyal Aftab\inst{1} \and
Jia Wu\inst{2}\and
Tanvir Alam\inst{3} \and
Zubair Shah\inst{3} \and 
Hazrat Ali\inst{3}
}

\authorrunning{U. Tariq et al.}
%
\institute{National University of Computer and Emerging Sciences, Karachi, Pakistan. 
\and
Department of Imaging Physics, MD ANDERSON Cancer Center, The University of Texas, Houston, Texas, USA.
 \and
College of Science and Engineering, Hamad Bin Khalifa University, Qatar Foundation, Doha, Qatar.\\
\email{K173810@nu.edu.pk, FRizwan@mdanderson.org, anaszafar98@gmail.com, danyalaftab97@gmail.com, JWu11@mdanderson.org, talam@hbku.edu.qa, zshah@hbku.edu.qa, haali2@hbku.edu.qa}
}
\maketitle              
\begin{abstract}
Generative models have been very successful over the years and have received significant attention for synthetic data generation. As deep learning models are getting more and more complex, they require large amounts of data to perform accurately. In medical image analysis, such generative models play a crucial role as the available data is limited due to challenges related to data privacy, lack of data diversity, or uneven data distributions. In this paper, we present a method to generate brain tumor MRI images using generative adversarial networks. We have utilized StyleGAN2 with ADA methodology to generate high-quality brain MRI with tumors while using a significantly smaller amount of training data when compared to the existing approaches. We use three pre-trained models for transfer learning. Results demonstrate that the proposed method can learn the distributions of brain tumors.  Furthermore, the model can generate high-quality synthetic brain MRI with a tumor that can limit the small sample size issues. The approach can addresses the limited data availability by generating realistic-looking brain MRI with tumors. The code is available at: ~\url{https://github.com/rizwanqureshi123/Brain-Tumor-Synthetic-Data}.
\end{abstract}
\section{Introduction}
Due to the advancements in computational power and a large amount of high-quality datasets, deep learning has become the state-of-the-art technology in computer vision, natural language processing, and others~\cite{ker2017deep}. Deep learning has also made remarkable progress in all areas of medical image analysis, including segmentation, detection, and classification~\cite{ker2017deep}. However, deep learning models are trained on large datasets, which may not be available in the medical domain due to privacy and ethical concerns~\cite{an2021hierarchical}. Medical experts find it difficult to publicize the majority of medical images without patients' consent. In addition, the public datasets are also small and lack expert annotations, thus, hindering their use for training deep neural networks. Furthermore, most of the available datasets might contain unbalanced classes that may hinder the performance of deep learning models and may not produce critical biological insights.

To overcome the problem of data unavailability, many researchers use generative models~\cite{goodfellow} to generate realistic synthetic images with diverse distributions for training complex deep learning models for medical analysis. Generative Adversarial Networks (GAN), a type of neural network, comprises two neural networks, one of which focuses on image production and the other on discrimination. The training of GAN involves a contest between the generator $G$ and the discriminator $D$. The discriminator $D$ is a binary classifier that determines if the data generated by $G$ belongs to the training set or not (real versus unreal). GANs can be used to create synthetic medical images, image captioning, and cross-modality image generation~\cite{yi2019generative,ali2022role}. Due to the adversarial training scheme's success in creating new image samples and utility in preventing domain shift, GANs have drawn great interest from the research community. However, a GAN with insufficient training data leads to over-fitting the discriminator. The feedback to the generator becomes meaningless, and the training starts to diverge~\cite{karras2020training}. A common approach to overcome over-fitting is data augmentation. For instance, training an image classifier by including images with rotation, noise, or scaling may increase the classifier's invariance to certain semantics-preserving distortions, which is a very desired quality. On the other hand, a GAN trained with comparable dataset augmentations learns to produce the augmented distribution~\cite{zhao2020feature}. 

Medical image analysis tasks such as brain tumor diagnosis ~\cite{sh_2016} are critical where one would wish for minimum error from a computer model. Brain tumor refers to excessive growth of cells in regions of the brain. An early diagnosis of a brain tumor increases the effectiveness of the treatment and hence, the survival rates. Early diagnosis of a brain tumor is necessary in order to treat it properly; otherwise, it might cause severe damage to the brain that can eventually be fatal. Magnetic Resonance Imaging (MRI) is the most popular way to generate brain scans and detect tumors in different regions of the brain. Many deep learning models~\cite{Ortega_2021} have been introduced recently to detect tumors in brain MRIs. However, progress is generally hindered by the lack of enough data. Traditional data augmentation methods, such as rotation, translation, mirror, and lightning, are not sufficient to generate a diverse, realistic dataset for brain tumor diagnosis. Synthetic images can be generated for this purpose which can address the problems associated with data acquisition, such as; privacy concerns, class imbalance, and small sample size.

Generative Adversarial Networks (GANs) have been very popular for generating realistic diversified datasets. In 2018 StyleGAN~\cite{Karras_Laine_Aila_2018} was proposed, with the main aim to improve the existing generator architecture $G$. StyleGAN mainly improved the existing architecture of the generator network in ProGAN~\cite{gao2019progan} for better performance and kept the discriminator $D$ network and loss functions constant. The latent code (\emph{z}) is transformed into an intermediate latent code (\emph{W}) prior to feeding it into the network. The synthesis network ($G$) is supervised by affine transforms through an adaptive instance that adds random noise maps to the space $W$ resulting in much entangled latent space. The proposed model is capable of generating realistic, high-quality images and offers control over the new style of the generated image.  

StyleGAN2~\cite{Karras_Laine_Aittala_Hellsten_Lehtinen_Aila_2019} architecture was presented to overcome issues present in the initial images generated by StyleGAN, such as blob and phase artifacts. Two causes were identified for the artifacts introduced in StyleGAN such as; fixed positions of eyes and nose and water droplet effects. Upon investigating the cause of common bob-like artifacts, it was observed that it was generated by the generator in response to an architectural design defect. A new design was proposed for the normalization used in the generator, which removed artifacts. 

In this paper, we used StyleGAN2 with adaptive discriminator augmentation (ADA)~\cite{Karras2020ada} for generating brain tumor MRI images of 512$\times$512 resolution while utilizing a significantly limited amount of training data when compared to the existing approaches. Our proposed approach effectively addresses the problem of data limitation by generating realistic brain MRI with tumor samples and can learn different data distributions from brain tumor raw images. The experiments are conducted on the brain tumor dataset. 
We utilized pre-trained models trained on FFHQ dataset~\cite{Karras_Laine_Aila_2018}, BreCaHaD dataset~\cite{Aksac_Demetrick_Ozyer_Alhajj_2019}, and AFHQ dataset~\cite{Kumari_Zhang_Shechtman_Zhu_2021}. The experimental results indicate that these models can generate superior quality superior MRI tumor samples that can be effectively utilized for medical analysis. 
The remaining paper is organized as: Section \ref{sec:literature} provides a review of the related literature. Section \ref{sec:proposed1} explains the methodology and the architecture. Section~\ref{sec:results} presents experiments, and Section~\ref{sec:results} presents the results and discussion for the synthesis of brain MRI having tumors. Finally, Section~\ref{sec:conclusion} concludes the paper. 
\section{Literature Review}
\label{sec:literature}
The SytleGAN architecture generates style images while controlling different high-level attributes of the images~\cite{Karras_Laine_Aila_2018}. The generator architecture in this research was designed in such a way that helps to control the image synthesizing process by learning on a constant input of 4x4x512 and on each subsequent layer based on latent code for adjusting the style of the image. When noise is provided as an input to the network, this combined effect helps segregate high-level attributes from stochastic variation
in generated images and allows for better style mixing and interpolation. The datasets used in the work are FFHQ~\cite{Choi_Uh_Yoo_Ha_2019}, LSUN~\cite{yu2015lsun} and CelebA-HQ~\cite{karras2017progressive}. The concept of intermediate latent space was used, which significantly affected how variational factors are represented in the network and could be disentangled. Two metrics, i.e., the perceptual path length and linear separability, were used to estimate the degree of latent space disentanglement.

StyleGAN2 was introduced in~\cite{Karras_Laine_Aittala_Hellsten_Lehtinen_Aila_2019} to address the characteristic artifacts and improve the output of StyleGAN. The first reason for the artifacts was the attempt of StyleGAN to evade a design flaw related to instance normalization used in AdaIN. The second type of artifact was related to progressive growth and was addressed in StyleGAN2 by changing the training method.
A method for mapping low-resolution medical images to high-resolution medical images using generative models is presented in~\cite{ahmad2022new}. In~\cite{Patashnik_Wu_Shechtman_Cohen-Or_Lischinski_2021}, considering the limitation of GANs to generate high-quality images for domains that have very little data, one of the very recent breakthroughs in generative modeling is a text-driven method that allows domain adaptation capability to the generator model for generating images across a multitude of domains. A text-driven method for out-of-domain image
synthesis is proposed. The domain shift was carried out by adjusting the generator's weights in the direction of images aligned with the driving text. The network architecture is dependent on StyleGAN2 and Contrastive-Language-Image-Pre-training (CLIP) \cite{Radford2016UnsupervisedRL,radford2021learning}.

CLIP model has been used for discovering semantically
rich and meaningful latent manipulations in order to generate images with styles defined through text based interface. In the first stage, an optimization task has been applied using CLIP-based loss to manipulate a latent input vector. In the next stage, a latent mapper for an optimized text-based manipulation given an input image has been used. Effectively, mapping the text-based inputs in the direction of StyleGAN style space results in effective text-based image manipulation. Motivated by the potential of the StayleGAN2 architecture to generate improved images of human faces, we use the StyleGAN2-ADA architecture to synthesize brain MRI, as explained in the following sections.

\begin{figure*}[!htb]
    \centering
    \includegraphics[width=12cm]{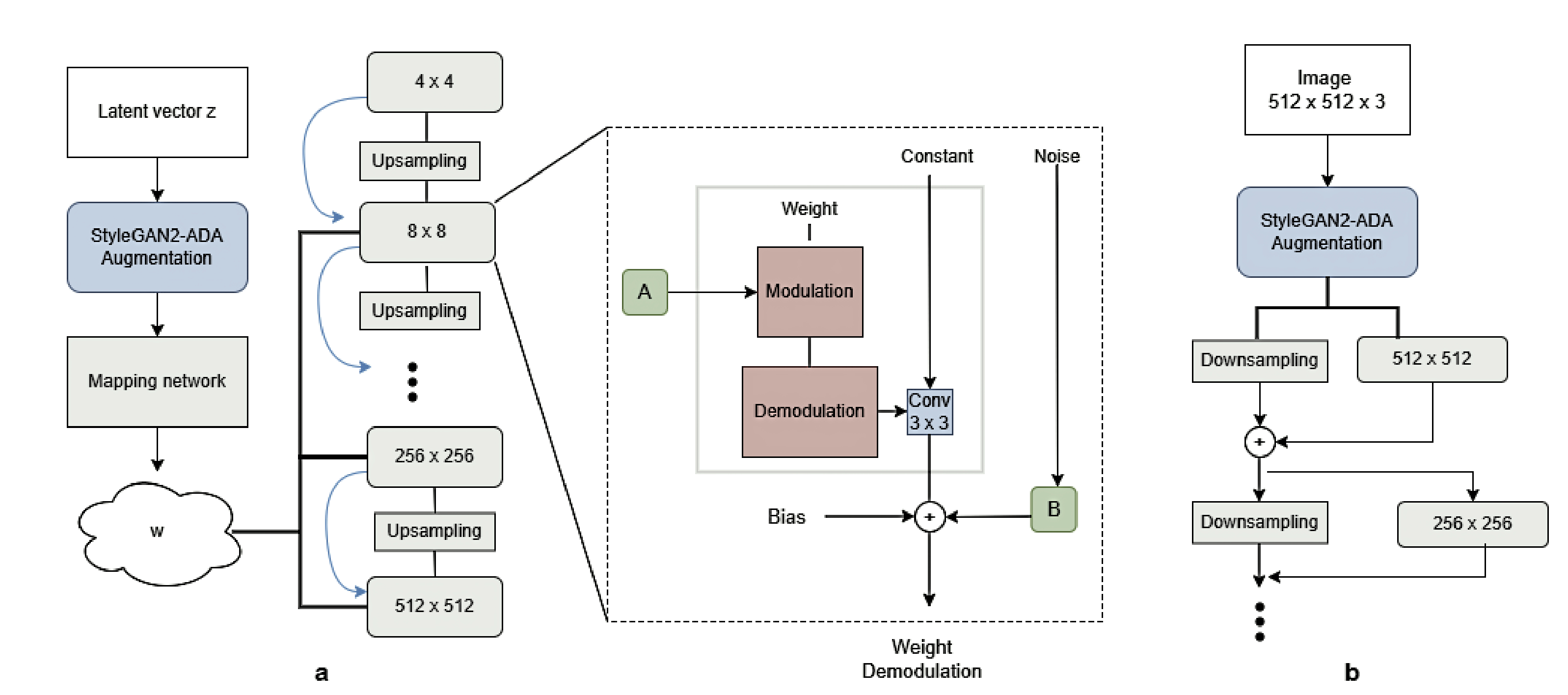}
    \caption{StyleGAN2-ADA (a) \textbf{Generator.} Based on the incoming style, the modulation scales each input feature map of the convolution, and the demodulation module is used to remove the droplet artifacts.
(b) \textbf{Discriminator.} After the input vectors of the components, StyleGAN2-ADA performs data augmentation.}
    \label{fig:proposed_model11}
\vspace{-5mm}
\end{figure*}
\section{Methodology}
\label{sec:proposed1}
We have utilized StyleGAN2 with ADA methodology to generate high-quality MRI brain tumor images while using a significantly limited amount of training data.
The proposed pipeline of StyleGANs with ADA is shown in Figure~\ref{fig:proposed_model11}. StyleGAN2~\cite{Karras_Laine_Aittala_Hellsten_Lehtinen_Aila_2019} introduced several changes in the architecture to overcome the issues in StyleGAN. Many viewers observed distinctive artifacts in StyleGAN images. Two key issues were identified in the output of StyleGAN, and changes were introduced in the architecture and the training method accordingly. Upon investigating the cause of common blob-like artifacts, it was observed that the blobs were generated in response to an architectural design defect. A new design was proposed for the normalization used in the generator, which helped in removing the artifacts. It was concluded that the artifacts related to progressive growing have been quite effective at stabilizing high-resolution GAN training. Overall, the following major improvements were made in the $G$ network considering issues in StyleGAN:
\begin{itemize}
  \item StyleGAN used a constant input \emph{c} as the model input directly, it was modified to input \emph{C} by adding noise and bias. 
  \item Noise and bias were moved outside the style block.
  \item Only the standard deviation value of every feature map was modified instead of modifying both the standard deviation and the mean values.
  \item Demodulation module was introduced to overcome the droplet artifacts.
\end{itemize}
\textbf{Weight Demodulation}: Similar to StyleGAN, StyleGAN2 makes use of a normalization technique to provide styles from the W vector using learning to transform $A$ into the source image. Here, the weight demodulation handles the droplet artifacts.

\textbf{Lazy Regularization}:
StyleGAN2 computes regularization terms once after 16 mini-batches compared to StyleGAN, which
computes both the main loss function and regularization for every mini-batch with heavy memory consumption and high computation cost. This change in approach is to compute the cost function, which has no major effects in terms of
model efficiency but speeds up the training.

\textbf{Path Length Regularization}:
Introducing path length regularization~\cite{Peebles_Zhu_Zhang_Torralba_Efros_Shechtman_2021} allows the same displacement in the latent space that would
produce the same magnitude change in the image space regardless of the value of the latent factor.

\textbf{Removing Progressive Growing}:
Progressive growth in StyleGAN causes phase artifacts (location preference for facial features).
StyleGAN2 overcomes the issue by using a different network design
based on skip connections similar to that of ResNet architectures.

\textbf{Adaptive Control Scheme}:
In order to have dynamic control over the augmentation strength parameter $p$ to avoid over-fitting, an adaptive control scheme ~\cite{Ma_Jin_Sohn_Paik_Chung_2019} has been used instead of manually tuning the augmentation strength. With the introduction of two heuristics to detect over-fitting in the discriminator, we are going to increase the magnitude of the augmentation to have dynamic scheduling. 

\begin{equation}
\label{eqn:rv}
r_v=E\left[D_{train}\right] - E\left[D_{validation}\right] /  E\left[D_{train}\right] - E\left[D_{generated}\right]
\end{equation}
\begin{equation}
\label{eqn:rt}
    r_t=E\left[sign(D_{train})\right]
\end{equation}

where $r_v$ is the first heuristic which refers to the validation set results relative to the training set and images generated given in Eq. (\ref{eqn:rv}). $r_t$ is the second heuristic that refers to the training set that generates positive discriminator outputs given in Eq. (\ref{eqn:rt}).
\section{Experiments}
\label{sec:results}
Style transfer learning mechanism is used for model training. Transfer learning~\cite{weiss2016survey} is used to reduce the training data by using the weights of the model already trained on a dataset~\cite{zhao2020feature,wang2020minegan,iqbal2018generative,wang2018transferring}.
\subsection{Datasets}
\begin{figure}[!htb]
    \centering
    \includegraphics[ height=3cm]{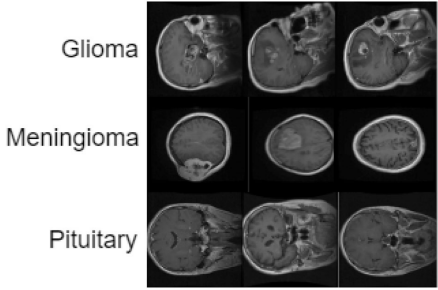}
    \caption{Brain tumor dataset sample images. Each row represents one type of tumor.}
    \label{data12}
\vspace{-5mm}
\end{figure}

We applied the models to the brain tumor dataset~\cite{Cheng_Huang_Cao_Yang_Yang_Yun_Wang_Feng_2015}~\cite{Singh}, as available via Kaggle \cite{Kaggle_data}. This dataset includes 154 brain MRI samples and contains 3064 T1-weighted images with high contrast consisting of three kinds of brain tumors which are classified as Glioma, Meningioma, and Pituitary Tumor, as shown in Figure~\ref{data12}.
\subsection{Implementation Details}

 We resize all training images to 512$\times$512 resolution. 
We used \emph{Google Colab Pro} platform for the training model as it allows access to faster GPUs which helps in speeding up the training. The model was trained on a Tesla P100 GPU with 25 GB RAM. For monitoring and managing GPU resources, NVIDIA System Management Interface (nvidia-smi) driver version $460.32.03$ and Cuda version 11.2 has been used for the management and monitoring of NVIDIA GPU devices. We converted all images into TFR records, enabling StyleGAN2 ADA to read data and improving the import pipeline's performance.
We utilized pre-trained models trained on FFHQ dataset~\cite{Karras_Laine_Aila_2018}, BreCaHaD dataset~\cite{Aksac_Demetrick_Ozyer_Alhajj_2019}, AFHQ \cite{Kumari_Zhang_Shechtman_Zhu_2021}.
\subsection{Pre-trained Models}
FFHQ512~\cite{Karras_Laine_Aila_2018} pre-trained model is trained on Flickr-Faces high-quality images (FFHQ) dataset. The FFHQ is an image dataset containing high-quality images of human faces. 
It offers 70,000 PNG images at 512$\times$512 resolution that display diverse ages, ethnicity, image backgrounds, and accessories like hats and eyeglasses.

BreCaHaD~\cite{Aksac_Demetrick_Ozyer_Alhajj_2019} pre-trained model is trained on a dataset consisting of 162 breast cancer histopathology images that are distributed into 1944 partially overlapping crops of 512$\times$512. The dataset is widely used by the biomedical and computer vision research community to evaluate and develop novel methods for tumor detection and diagnosis of cancerous regions in breast cancer histopathology images.

Animal FacesHQ~\cite{Choi_Uh_Yoo_Ha_2019} (AFHQ) pre-trained model is trained on a dataset of 15,000 high-quality animal face images at 512$\times$512 resolution in three domains of cat, dog, and wildlife, with 5000 images per domain.
AFHQ sets a more challenging image-to-image translation problem by having three domains and diverse images of various breeds. The images are vertically and horizontally aligned. The low-quality images were manually discarded. We used weights from the AFHQ (Cat) and AFHQ (Wild) pre-trained models. 

\subsection{Evaluation Metrics}
Fréchet Inception Distance (FID)~\cite{Nunn_Khadivi_Samavi_2021} is a metric for quantifying the distance between two distributions of images $P_r$ and $P_g$ where $P_r$ is the probability distribution of real images, and $P_g$ is the probability distribution of generated images. It is used to evaluate the quality of generated images and the performance of GANs. FID is defined as:
\vspace{1mm}
\begin{equation}
\operatorname{FID}\left(\mu_r, \Sigma_r, \mu_g, \Sigma_g\right)=\left\|\mu_r-\mu_g\right\|_2^2+\operatorname{Tr}\left(\Sigma_r+\Sigma_g-2\left(\Sigma_r \Sigma_g\right)^{1 / 2}\right)
\end{equation}
where ($\mu_r$, $\Sigma_r$) and ($\mu_g,\Sigma_g$) denote the mean vectors and covariance matrices of the Gaussian approximations for real and generated samples, respectively. The lower the FID value, the better the generated image quality.
Kernel Inception Distance (KID)~\cite{Knop_Mazur_Spurek_Tabor_Podolak_2022} is a metric which measures the dissimilarity between two probability distributions $P_r$ and $P_g$ using samples drawn independently from each distribution. KID is defined to be
the squared maximum mean discrepancy (MMD) between the Inception features of real and generated
images. A cubic polynomial kernel is used to map the real and generated images from the feature
space of the Inception network, which is defined as:
\begin{equation}
KID(x,y) = \left (\frac{1}{d}{x^T}y + 1  \right ) \textsuperscript{3}
\vspace{-5mm}
\end{equation}
 
 \begin{figure}[!htb]
    \centering    
    \includegraphics[height=4cm]{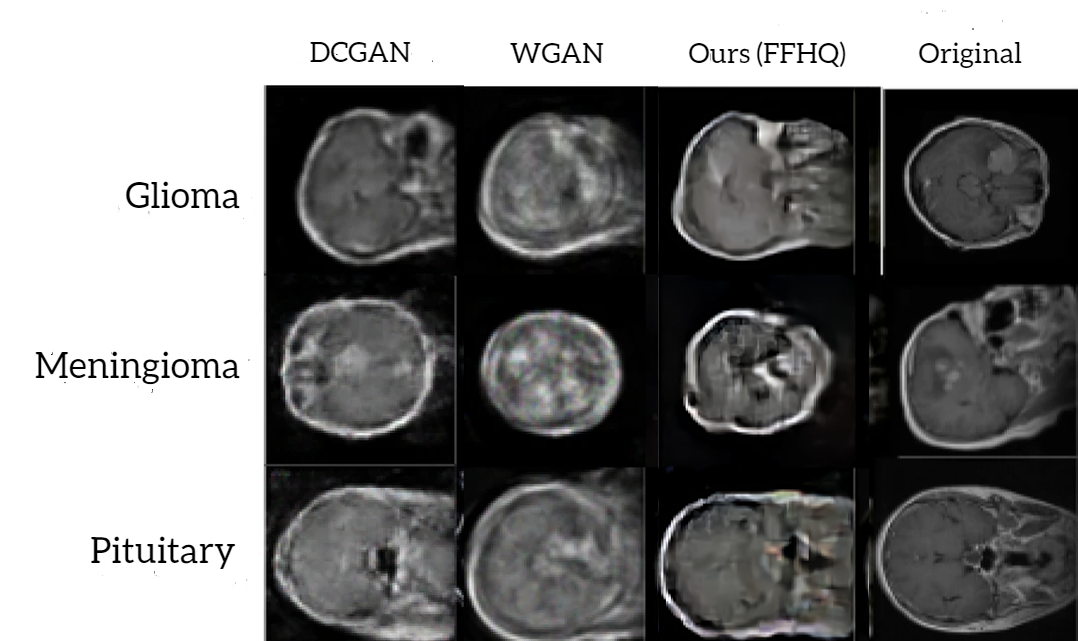}
    \caption{Comparative analysis of generated brain tumor MRI samples using DCGAN~\cite{Radford2016UnsupervisedRL}, WGAN~\cite{wgan}, FFHQ (Ours), and the original sample. Each row corresponds to images from three different classes, namely; Glioma, Meningioma, and Pituitary.}
    \label{fig:compare}
\vspace{-10mm}
\end{figure}


\begin{figure}[!thb]
     \centering
     \begin{subfigure}[b]{0.45\textwidth}
         \centering
         \includegraphics[width=\textwidth]{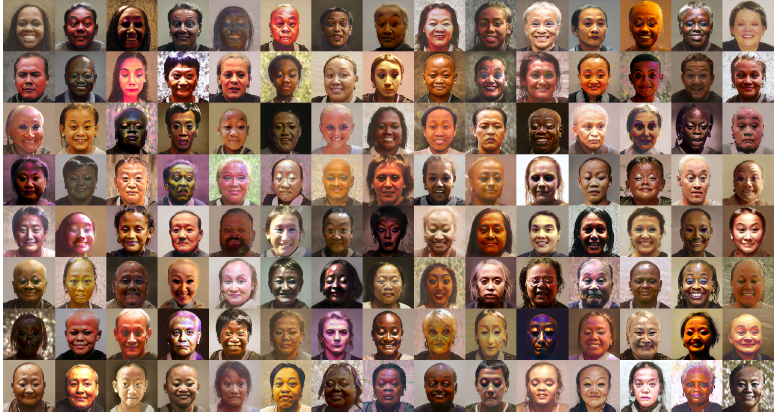}
         \caption{Initial synthetic images using FFHQ}
         \label{fig:a}
     \end{subfigure}
     \begin{subfigure}[b]{0.45\textwidth}
         \centering
         \includegraphics[width=\textwidth]{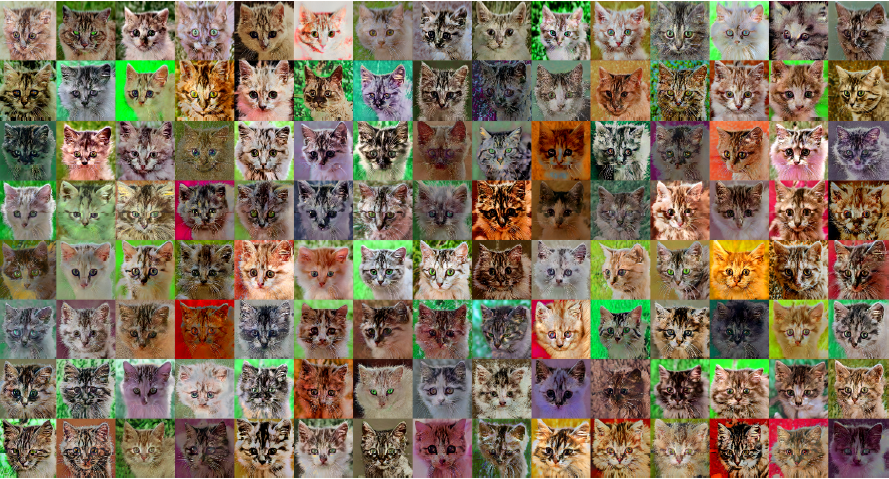}
         \caption{Initial synthetic images using AFHQ (Cat)}
         \label{fig:b}
     \end{subfigure}
     \hfill
     \begin{subfigure}[b]{0.45\textwidth}
         \centering
         \includegraphics[width=\textwidth]{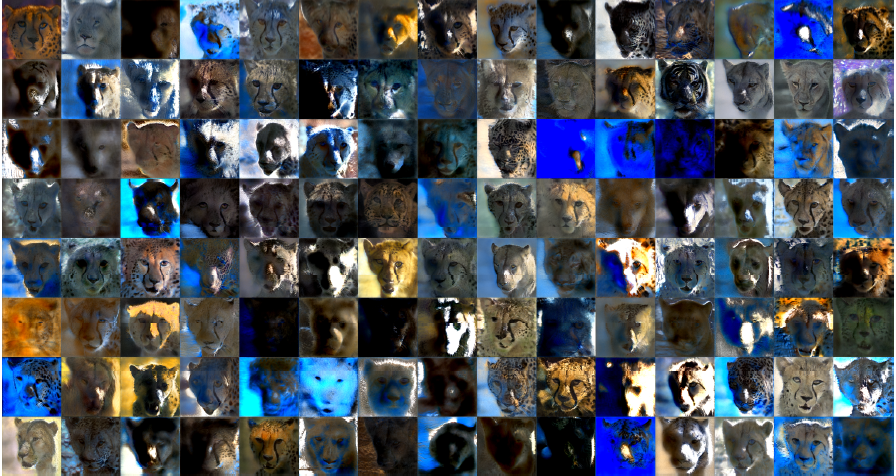}
         \caption{Initial synthetic images using AFHQ (Wild)}
         \label{fig:c}
     \end{subfigure}
     \begin{subfigure}[b]{0.45\textwidth}
         \centering
         \includegraphics[width=\textwidth]{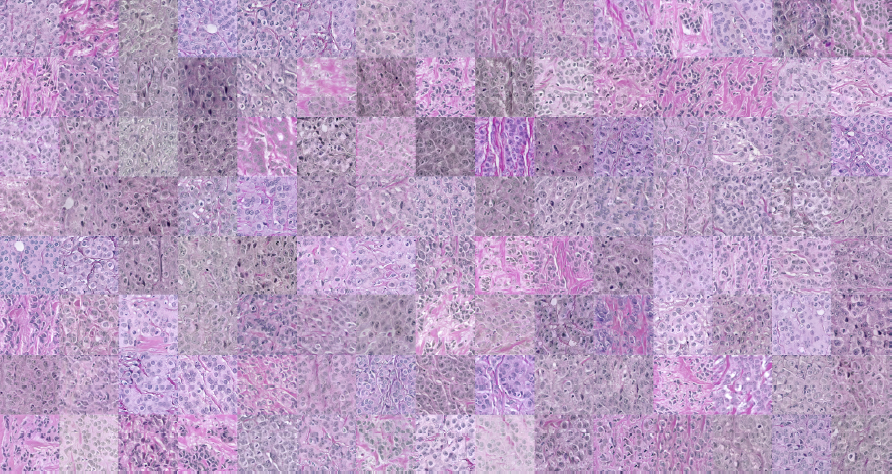}
         \caption{Initial synthetic images using BreCaHaD (Histopathology)}
         \label{fig:d}
     \end{subfigure}
        \caption{Samples of initially generated images. Results show a visualization of the weights of the StyleGAN model trained on FFHQ, AFHQ (Cat), AFHQ (Wild), and BreCaHad images.}
        \label{fig:initialimages}
\vspace{-7mm}
\end{figure}

\begin{figure}[!htb]
     \centering
     \begin{subfigure}[b]{0.7\textwidth}
         \centering
         \includegraphics[width=\textwidth]{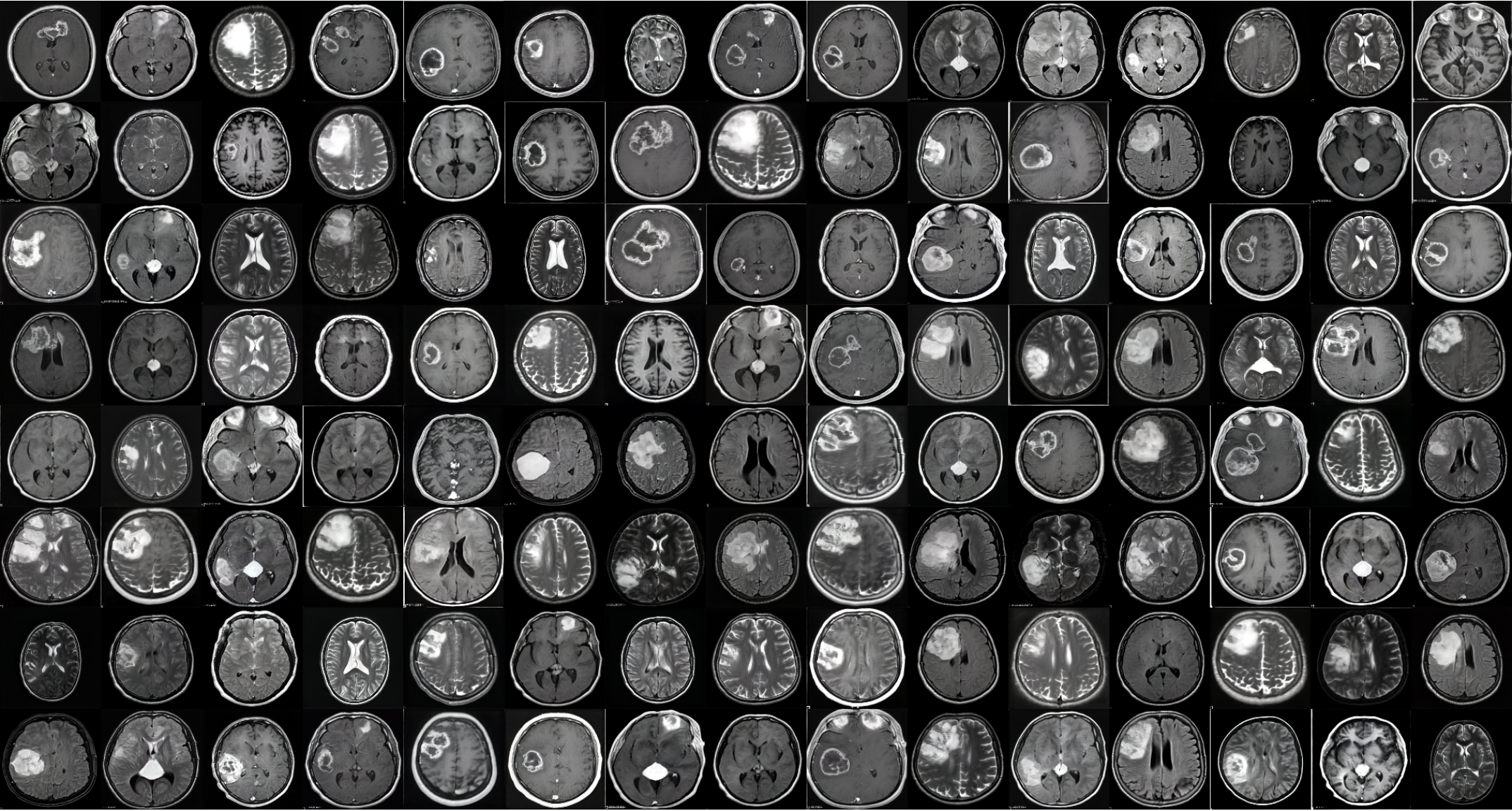}
         \caption{Synthetic MRI brain tumor images on best FID and KID for the FFHQ pre-trained model.}
         \label{FFHQB}
     \end{subfigure}
     \hfill
     \begin{subfigure}[b]{0.7\textwidth}
         \centering
         \includegraphics[width=\textwidth]{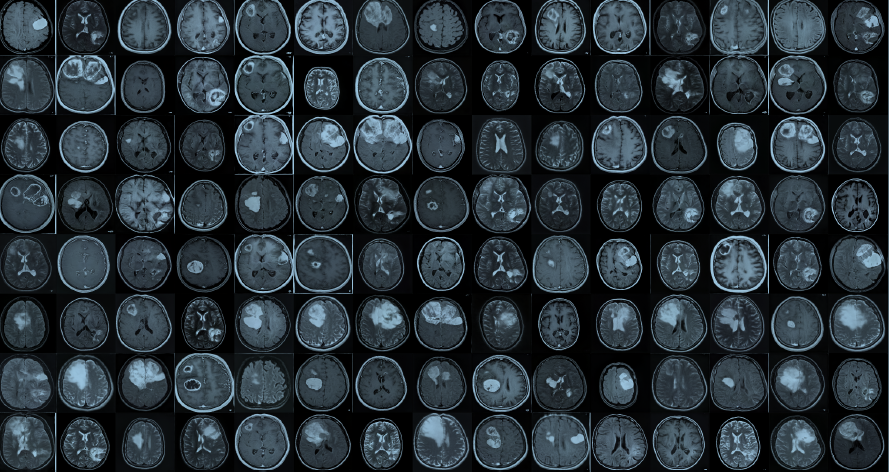}
         \caption{Synthetic MRI brain tumor images on best FID and KID values for the AFHQ (Cat) pre-trained model.}
         \label{AFHQB}
     \end{subfigure}
     \hfill
     \begin{subfigure}[b]{0.7\textwidth}
         \centering
         \includegraphics[width=\textwidth]{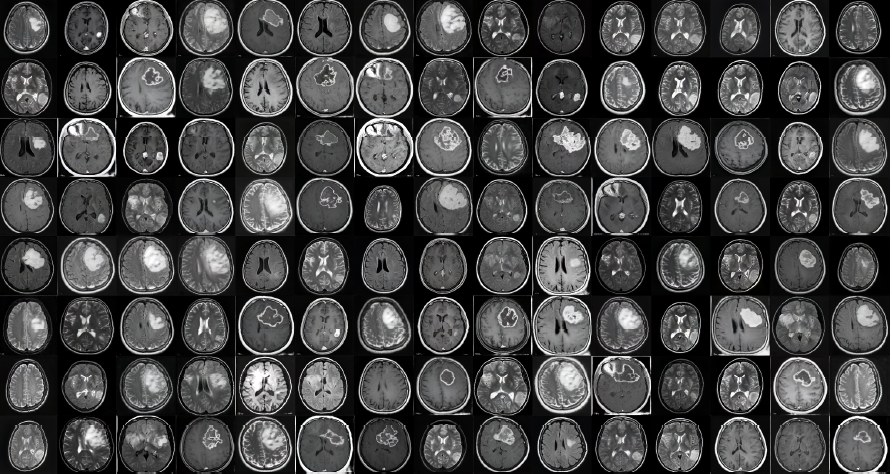}
         \caption{Synthetic MRI brain tumor images on best FID and KID values for the AFHQ (Wild) pre-trained model.}
         \label{AFHQWB}
     \end{subfigure}
     \begin{subfigure}[b]{0.7\textwidth}
         \centering
         \includegraphics[width=\textwidth]{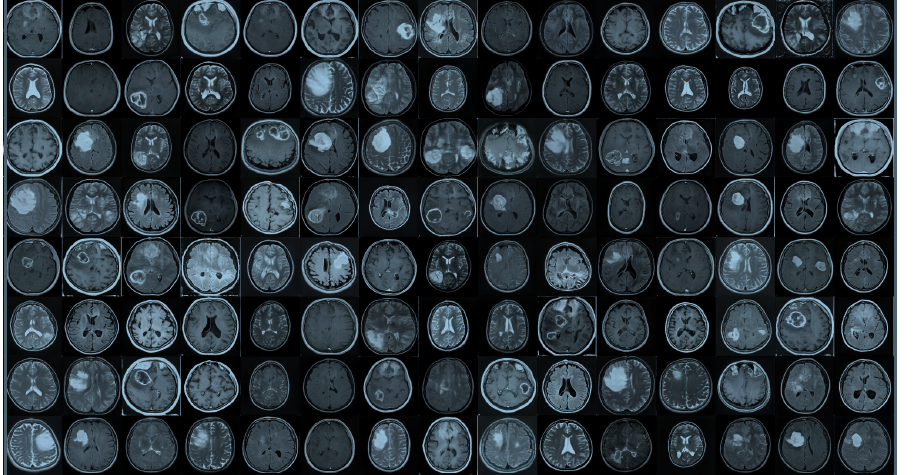}
         \caption{Synthetic MRI brain tumor images on best FID and KID values for the BreCaHaD pre-trained model.}
         \label{fig:five over x}
     \end{subfigure}
        \caption{Synthetic images generated for the best FID and KID values.}
        \label{fig:bestfidkid}
\end{figure}
\vspace{-2mm}
\section{Results and Discussion}
FID is a commonly used metric for assessing the quality of the images generated by a model. However, FID is prone to be dominated by the inherent bias when the number of real images is not large enough. Hence, we used KID as an additional metric for evaluating our model performance. 
The trend for FID and KID using different pre-trained models is shown in Figure~\ref{pre14}. FID and KID values were recorded on every ten tick intervals for FFHQ, BreCaHaD, and AFHQ models, where tick interval refers to the number of iterations after the training snapshot has been taken. The results are summarized in Table~\ref{tab:pre_trained_model}.
 \par The trend indicates a decrease in FID and KID values as tick intervals increase for FFHQ and AFHQ (Cat) models. For BreCaHaD and AFHQ (Wild) models, a decrease can be observed from 0-30 tick intervals. After that, an increase can be seen for both FID and KID values. Amongst the models evaluated, the BrecaHaD model had the worse performance, having the highest FID and lowest KID values. 
 \par Qualitative results of initially synthetically generated brain tumor images by different models are shown in Figure~\ref{fig:initialimages}.  
Using the best FID and KID of the pre-trained models, the brain MRI images generated by transfer learning are shown in Figure~\ref{fig:bestfidkid}. By analyzing our results, we find that FFHQ gives the lowest
FID of 58.1097 and KID of 0.00862, and generates better quality images
when compared with other pre-trained models. Figure~\ref{fig:compare} shows a comparison of the images generated using DCGAN~\cite{Radford2016UnsupervisedRL}, WGAN~\cite{wgan} and Ours (FFHQ) model using the brain tumor dataset. The results indicate Ours (FFHQ) generates better quality images when compared with the other GAN models.
\begin{table}[!htb]
\begin{center}
\caption{Results}
\label{tab:pre_trained_model}
\begin{tabular}{|p{3cm}|p{2cm}|p{3cm}|}
\hline
Pre-trained models & FID & KID\\
\hline
FFHQ & 58.1097 & 0.00862692\\
AFHQ Cat & 60.9486 & 0.01049849\\
BreCaHaD & 67.5336 & 0.02081763\\
AFHQ Wild & 59.7498 & 0.0109629\\
\hline
\end{tabular}
\end{center}
\vspace{-8mm}
\end{table}
\section{Conclusion and Future Work}
\label{sec:conclusion}
In this work, we presented a useful application of Adaptive StyleGANS for synthetic brain MRI images. Our results show that high-quality realistic MRI brain tumor images can be generated using pre-trained GAN models. 
By analyzing our results, we find that FFHQ gives the lowest FID and KID and generates better quality images when compared with other pre-trained models used in this research. This work will motivate other researchers to leverage the potential of StyleGAN in many applied domains of medical imaging research. For example, the models can be explored for modeling to detect the presence of tumors in body parts, perform tissue segmentation when training largely suffers due to the unavailability of high-quality data, and cross-modality medical image generation. The future work of this research is to explore the use of StyleGAN2-based architectures for the synthesis of high-quality medical images of other modalities such as Computed Tomography and histopathology images.
It would be interesting to evaluate the model performance with other smaller medical imaging datasets. Similarly, an interesting direction is to explore the use of StyleGAN2 with StyleCLIP ~\cite{Patashnik_Wu_Shechtman_Cohen-Or_Lischinski_2021} for generating medical images from the text description. 

%
%
%







\end{document}